\documentclass[hidelinks, journal, a4paper, compsoc]{IEEEtran}

\usepackage[latin1]{inputenc}
\usepackage[spanish, english]{babel}

\usepackage{subcaption}

\usepackage{url}

\usepackage{graphics,graphicx} 
\usepackage{wrapfig} 
\usepackage[rflt]{floatflt} 
\usepackage{graphpap}	
\usepackage{graphicx}

\usepackage{tabularx} 
\usepackage{multirow}

\usepackage{textcomp}
\usepackage{listings}

\usepackage{xcolor}

\colorlet{punct}{red!60!black}
\definecolor{background}{HTML}{FDFDFD}
\definecolor{delim}{RGB}{20,105,176}
\colorlet{numb}{magenta!60!black}

\lstdefinelanguage{json}{
	basicstyle=\normalfont\ttfamily,
	numbers=left,
	numberstyle=\scriptsize,
	stepnumber=1,
	numbersep=8pt,
	showstringspaces=false,
	breaklines=true,
	frame=single,
	backgroundcolor=\color{background},
	literate=
	*{0}{{{\color{numb}0}}}{1}
	{1}{{{\color{numb}1}}}{1}
	{2}{{{\color{numb}2}}}{1}
	{3}{{{\color{numb}3}}}{1}
	{4}{{{\color{numb}4}}}{1}
	{5}{{{\color{numb}5}}}{1}
	{6}{{{\color{numb}6}}}{1}
	{7}{{{\color{numb}7}}}{1}
	{8}{{{\color{numb}8}}}{1}
	{9}{{{\color{numb}9}}}{1}
	{:}{{{\color{punct}{:}}}}{1}
	{,}{{{\color{punct}{,}}}}{1}
	{\{}{{{\color{delim}{\{}}}}{1}
	{\}}{{{\color{delim}{\}}}}}{1}
	{[}{{{\color{delim}{[}}}}{1}
	{]}{{{\color{delim}{]}}}}{1},
}

\usepackage{adjustbox}
\usepackage{longtable}

\usepackage[english, plain]{fancyref} 
\fancyrefchangeprefix{\fancyreftablabelprefix}{table}
\newcommand*{\fancyrefpartlabelprefix}{part}
\fancyrefchangeprefix{\fancyrefpartlabelprefix}{part}

\usepackage{enumitem}

\usepackage[parfill]{parskip}

\usepackage{booktabs}

\newlength{\tpw}
\newcommand\asta{1.3} 

\newcolumntype{P}[1]{>{\RaggedRight\hspace{0pt}}p{#1}}
\newcolumntype{M}[1]{>{\RaggedRight\hspace{0pt}}m{#1}}

\newcolumntype{Q}[1]{>{\centering\let\newline\\\arraybackslash\hspace{0pt}}p{#1}}
\newcolumntype{N}[1]{>{\centering\let\newline\\\arraybackslash\hspace{0pt}}m{#1}}

\newcolumntype{R}[1]{>{\RaggedLeft\hspace{0pt}}p{#1}}
\newcolumntype{O}[1]{>{\RaggedLeft\hspace{0pt}}m{#1}}

\author{
	\IEEEauthorblockN{Sergio~Fortes\IEEEauthorrefmark{1}, \textit{IEEE Member}, 
		David~Palacios\IEEEauthorrefmark{1}, Inmaculada~Serrano\IEEEauthorrefmark{2}, Raquel~Barco\IEEEauthorrefmark{1}, }
	\vspace{\baselineskip} 
	\fontsize{8pt}{10pt}\selectfont
	
	\IEEEauthorblockA{\IEEEauthorrefmark{1}
		Universidad de Málaga, Andalucía Tech, Departamento de Ingeniería de Comunicaciones, Campus de Teatinos s/n, 29071 Málaga, España
		\\\{sfr, rbm, dpc\}@ic.uma.es}
	
	\vspace{0.5\baselineskip} 
	
	\IEEEauthorblockA{\IEEEauthorrefmark{2}Ericsson, Málaga 29590, Spain
		\\inmaculada.serrano@ericsson.com}
}

\usepackage{bm}
\usepackage{textpos}

\usepackage{hyperref}
\hypersetup{
    colorlinks=true,
    citecolor=black,
    linkcolor=black,     
    urlcolor=cyan,
}

\begin{document}
	
	\title{Applying Social Event Data for the Management of Cellular Networks}
	
	\markboth{First submitted to "IEEE Communications Magazine" in June 17, 2017. Accepted version in Jan 7, 2018}%
	{}
	
	\maketitle


\begin{textblock*}{\textwidth}(0cm, -10.2cm)
\small {\color{red}
 This is the author's version of an article that has been accepted for publication in IEEE Communications Magazine. Changes were made to this version by the publisher prior to publication. The final published article is available at \url{https://doi.org/10.1109/MCOM.2018.1700580} }
\end{textblock*}

\begin{textblock*}{\textwidth}(0cm, 17 cm)
\small {\color{red}\textcopyright 2020 IEEE. Personal use of this material is permitted.  Permission from IEEE must be obtained for all other uses, in any current or future media, including reprinting/republishing this material for advertising or promotional purposes, creating new collective works, for resale or redistribution to servers or lists, or reuse of any copyrighted component of this work in other works.}
 
\end{textblock*}

\hypersetup{
    colorlinks=true,
    citecolor=black,
    linkcolor=black,     
    urlcolor=black,
}


	\begin{abstract}
		Internet provides a growing variety of social data sources: calendars, event aggregators, social networks, browsers, etc. Also, the mechanisms to gather information from these sources, such as web services, semantic web and big data techniques have become more accessible and efficient. This allows a detailed prediction of the main expected events and their associated crowds. Due to the increasing requirements for service provision, particularly in urban areas, having information on those events would be extremely useful for Operations, Administration and Maintenance (OAM) tasks, since the social events largely affect the cellular network performance. Therefore, this paper presents a framework for the automatic acquisition and processing of social data, as well as their association with network elements (NEs) and their performance. The main functionalities of this system, which have been devised to directly work in real networks, are defined and developed. Different OAM applications of the proposed approach are analyzed and the system is evaluated in a real deployment. 
	\end{abstract}
	
	

	\section{Introduction}
	
	Cellular networks complexity and extent is continuously growing. Accordingly, the cost of their OAM tasks increases pairwise. In this field, the concept of Self-Organizing Network (SON) was defied with the objective of automating some of the main tasks of OAM, including self-configuration, on the plug\&play capabilities, planning and replanning of NEs, such as the cell stations; self-optimization, on the automatic change of network parameters to cope with variations in the service demands and conditions (e.g., user distributions, capacity demand); and self-healing on the detection, diagnosis and prediction of network failures and their compensation and prevention.
	
	Current techniques associated with these OAM tasks are based on the own network monitoring information, such as counters, alarms and key performance indicators (KPI) \cite{DBLP:journals/cm/BarcoLL12:framework}. In cellular networks, human crowds or events (e.g., concerts, sport matches, demonstrations, etc.) can cause a huge impact on network performance. However, in current management procedures, these social events are generally not considered in an automatic manner. Commonly, only the most relevant/popular periodic events (e.g., fairs) are taking into account. Temporal solutions to cope with their associated increase in demand are implemented, for example, by the deployment of temporal base stations. 
	
	On the one hand, the list of events that can impact the cellular network performance is too large for their complete ``manual'' identification and follow-up. Besides, many of the events are non-periodical and difficult to identify a priori. These might only be identified by manual checks on social media (newspaper, event agendas) after a degradation in the service provision has occurred. In this way, finding and relating events with cell performance is a very time-consuming task and requires an important knowledge of the area.

	On the other hand, social events and their details can be predicted by automatically gathering data from Internet sources: particularly, event-listings coming from calendars, event aggregators, web services and databases. Additionally, crowds can be predicted based on social networks and activity sources \cite{2017:Chen:Crowd}\cite{Panagiotou2016}. 
	
	
	However, previous approaches in the use of non-cellular information in the OAM /SON activities, have mainly focused on applying user equipment (UE) related data, such as their location \cite{Fortes:2015:ContextArchit}\cite{Fortes:2015:ContextSH}. The use of information from social events has been, up to our knowledge in the field, disregarded by existing cellular network management procedures. Until recent times, extensive social databases (DBs) were very limited, as well as the acquisition procedures to gather this information. Also, the reluctance of operators to include information outside their control has favored this situation.
	
	\begin{figure*}[h!] 
		\centering
		\includegraphics[width=2\columnwidth]{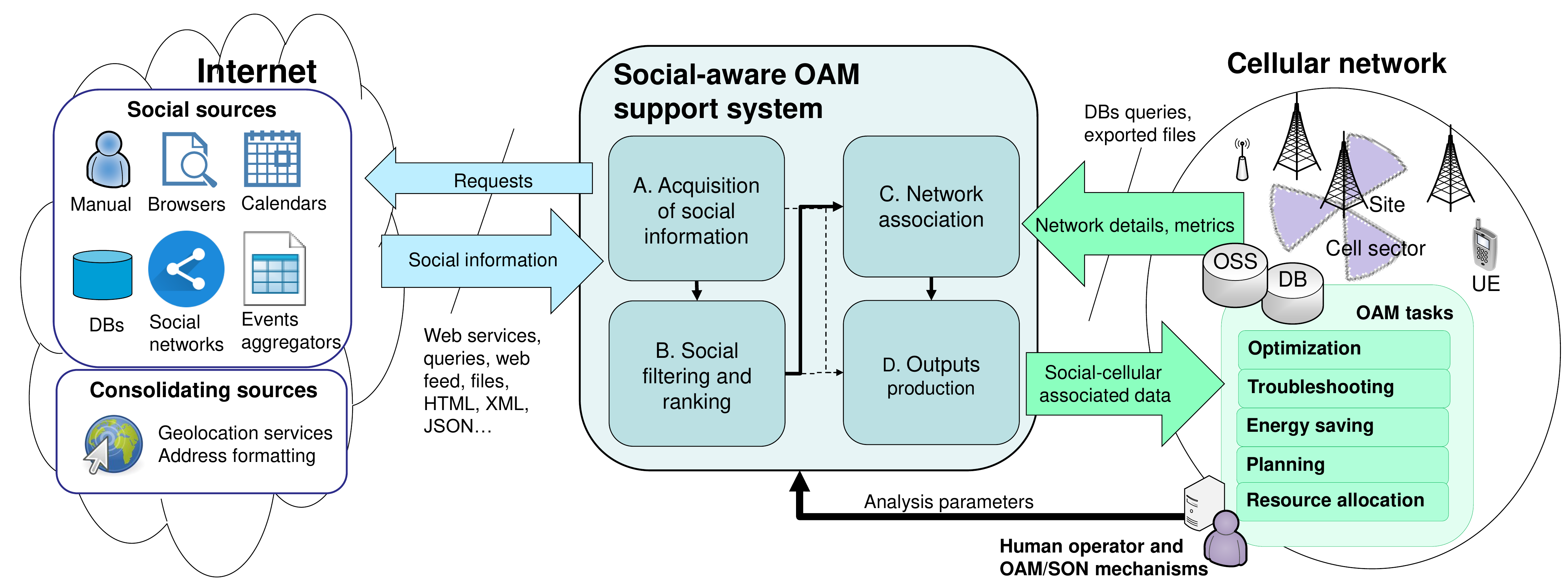}
		\caption{High level diagram of the proposed framework for the use of social events data in OAM tasks.}
		\label{fig:Fig_1}
	\end{figure*}

	Most previous approaches make use of network-related information for social applications, and not vice versa. Multiple works have followed this concept, making use of cellular-network provided data (such as cellular-based location) and/or the cellular communications to support social network services such as event recommendations. For example, in \cite{zheng2015mobile} event recommendations for mobile terminals are based on their location and preferences is proposed, when the event information comes from external data sources. Although it provides some general concepts on social events gathering, it focuses solely in supporting end-user services. It also refers only to social-network sources, which are commonly unreliable, and often only provides derivative or unformatted data. The work in \cite{cavanaugh2014real} presents a system for real time event notification by the UE to another party (e.g., a social network), sending at the same time mobile terminal information. 
	
	A system for prediction of pedestrian flows in urban areas is defined in \cite{2016:Zhang:FlowUMTS}, focusing on the analysis of cellular network data to perform the analysis. In the same line, other works have focused in the analysis of mobile data for smart-city applications. For example, looking at the identification of features in the urban ecology (e.g., social interaction)\cite{Cici:2015:DCP:2746285.2746292}. 
	
	Existing solutions associated with the capture of non-network data are applied only to very specific non-OAM tasks. In \cite{HAYES_AMEND}, an adaptable handover algorithm for vehicular environments is developed. This makes use of the conditions of the environment, specifically the foliage density and wind speed, information which might be gathered automatically from Internet data sources. Other attempts of using social-media, like Twitter, have been also applied only to non-OAM activities, as geolocation and estimating mobile traffic demand areas \cite{2016:Yang:Twitter}. The work in \cite{CiciCells}, establishes a predictive mechanism of the amount of traffic cell-to-cell based on previous cellular records and aiming at possible smart city applications. Also, a hint is provided on the impact that an unexpected social event may cause to the network, showing how the search for the identification of that events is typically performed manually (e.g. by web search). 
	
	Therefore, existing solutions do not provide the necessary means for the application of social events for the OAM tasks of cellular networks. Given the availability of sources that can provide detailed information about social events, and the interest of using this in the management of cellular networks, the paper proposes a novel general framework for the automatic acquisition of events data, its association with cellular network information and its application in OAM/SON tasks. The proposed developments have been included in a patent application that has been filed on May 1, 2017 (application number PCT/EP2017/060312).
	
	This paper is organized as follows. \Fref{sec:ProposedSolution} presents the variables to be considered in social-aware OAM analyses. In \Fref{sec:methodology}, the proposed social-aware OAM support system is defined, describing in detail its inputs and each of its functional blocks. Then, in \Fref{sec:Evaluation}, the algorithm is evaluated by using a collection of faults from a live network. Finally, \Fref{sec:Conclusion} presents the main conclusions. 

	\section{Inputs for social-aware network analyses} \label{sec:ProposedSolution}
	
	The proposed framework for the acquisition and application of social events in OAM/SON tasks follows the scheme shown in \Fref{fig:Fig_1}, whose details are described in the following subsections. 
	
	Three main types of data sources are considered for an integrated social-aware support to OAM: the \textit{social sources}, the \textit{cellular network} and \textit{human operator and SON mechanisms}. 
	
	Different \textit{social sources} can provide events information: venue and city-wide calendars, browser results, social networks, open DBs, manual-input data and event aggregators. The framework can simultaneously get information from various of these sources, even with different access procedures/formats: DB queries, web service APIs, data files, etc. Event aggregators (e.g. EventBrite, Eventful, Last.fm, etc.) have a special interest, as they commonly allow to obtain large number of events for any relevant area and using easily to automate API calls. A very detailed description and procedures of these kind of services can be found in the work in \cite{Becker:2012:ICP:2124295.2124360}.

	From these sources, semantic information of the different events is obtained, such as START\_TIME, STOP\_TIME, LAT (latitude), LON (longitude), TYPE (e.g., musical, parade, sport, etc.), VENUE, ADDRESS and POPULARITY. The obtained events data typically contains some inconsistencies in the precision of the geographical coordinates, address formatting/completeness, etc. Consolidating sources (such as geocoding solutions \cite{GeocodingDisparity}) are applied to correct these.

	Regarding the information from the cellular networks, the details of the \textit{sites} and cell sectors are of main importance. Network metrics, such as counters, KPIs, alarms, traces, contextualized indicators \cite{Fortes:2015:ContextIndicators} or any other network parameter or measurement variable in time are essential as well. This information can be gathered directly by query to the OAM databases of the cellular network, or from replicated copies or exported files coming from them. 
	  
	The human operator or external OAM/SON mechanisms have to establish the scope for each analysis or execution of the system. For this, both the temporal and geographical scopes should be defined (e.g., the coordinates of the location and the dates of interest). Also, the specific social and network sources and the procedures to acquire and process their data should be configurable, although they would be expected to remain common to most analyses. 
	
	Given that all the framework inputs can be accessed remotely and by common procedures (DBs queries, API calls), there would be a high level of freedom in the selection of its coding language and deployment platform.  

	\section{Social-aware OAM support system}\label{sec:methodology} 
	
\begin{figure}[ht] 
		\centering
		\includegraphics[width=1\columnwidth]{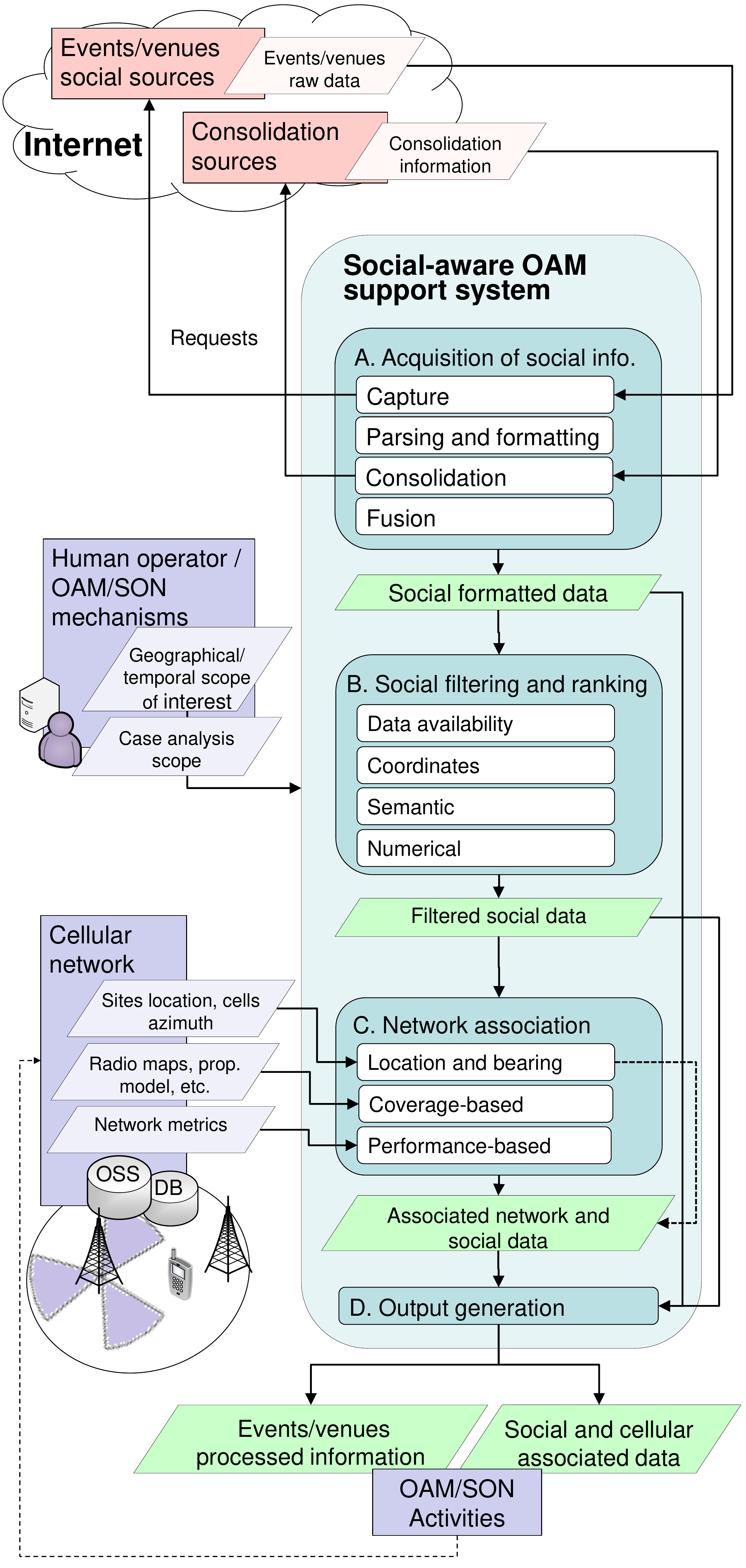}
		\caption{Detailed functional steps of the proposed system.}
		\label{fig:Fig_2}
	\end{figure}

	\Fref{fig:Fig_2} presents the proposed system, which is divided into four logical modules: acquisition of social information, social filtering and ranking,  cellular network association and outputs production. Each block is detailed in the subsections below. 
	
	\subsection{Acquisition of social information} 
	
	This module contains the different steps dedicated to acquiring and consolidating the social information from different sources. In addition, it formats the data to be passed to further stages of the system. 
	
	
	To do so, first, the \textit{captured information from social sources} is needed. To access the social sources, different procedures and parameters might be applied. These parameters include the type of connection, the address, the call format, etc. Multiple requests to the sources might be produced, demanding the available information for the temporal and geographical scopes of interest defined by the human operator or the SON external mechanisms.

	
	The different information replies are \textit{parsed} (interpreted from their original format, commonly string-based) and then \textit{formatted} and stored following a common internal data structure. This internal format unifies the identification of the different variables of the data of events or venues. For example, for an event, these variables include the NAME, START\_TIME, END\_TIME, LAT, LON, VENUE, etc. In this way, to refer to the value of one of these variables for a specific event, the nomenclature used will be EVENT.$<$VARIABLE$>$.For example EVENT\_A.START\_TIME.

	The \textit{details consolidation} step is in charge of completing and correcting the data based on operator or third party consolidating sources/services for address completion, geolocation and venue details. Also, this step could serve to include additional information, such as counting the number of appearances of the events in social media. 

	Besides, data for the same event could be retrieved multiple times from different sources. Also, the same elements can sometimes appear multiple times in the same source. In order to avoid repetitions, a \textit{sources fusion} process is dedicated to eliminate the duplicities by discarding or joining elements with the same or very close names and dates than already existent ones. This kind of analysis is performed through string similarity scores for specific fields or implemented through more complex semantic matchings. 
	
	\subsection{Social filtering and ranking}
	
	This module applies different filtering and ranking procedures to reduce the number of elements to be processed in later stages, as well as to classify the events depending on their expected relevance. At this stage, the analysis is purely based on social data (e.g., considering the event popularity or type) and not on cellular information.

	Even after the details consolidation and source fusion steps, some events might be still missing in some fields (e.g., START\_TIME, LAT, LON). The lack of values for such variables might make the social data useless for the system purposes, being appropriate to delete them or minimize their ranking based on \textit{data availability}.   
	
	Also, even if the requests to the social sources are defined for a certain area, some data outside the intended geographical scope might be returned, especially if it was defined categorically (e.g., indicating a city name), so additional \textit{geographical filtering} is needed. Typically, in the acquisition phase, the geographical scope would be provided in a categorical way [CITY, REGION, COUNTRY] to assure the acquisition of all the data in the scope. Meanwhile, for the coordinates-based filtering, once the geographical coordinates of the events have been consolidated by additional requests to consolidation sources, coordinate-based limits would be applied, e.g., by a square area of interest (providing LAT\_MAX, LAT\_MIN, LON\_MIN, LON\_MAX limits) or circular (defined by the coordinates of a CENTRAL\_POINT and a RADIUS).

	Afterwards, \textit{semantic processing} is performed analyzing different aspects of the acquired data of the events, typically:
	
	\begin{itemize}
		\item \textit{Key terms and category}: based on the presence of different words in their data, it can be inferred that certain elements have little relevance in the network (as they will not generate large crowds or high demand). For example, the elements that include in its venue terms like \{``bar'', ``pub'', ``tavern''\} might be discarded, as they typically imply small venues.

		\item \textit{Address and dates}: Complementary to the filtering of the previous step, the retrieved address details (street, city, etc.) and date can be used for filtering out elements outside the intended temporal or geographical scopes.

	\end{itemize}
	
	The posterior \textit{numerical processing} submodule is defined to adopt further ranking and filtering processes based on the variables which have a quantitative nature, such as the number of sold tickets, the capacity of the venue or other popularity scores. 
	
	\subsection{Network association}
	
	This step focuses on identifying the NEs (particularly sites and cells) and network metrics that had been or are most likely to be impacted in the future by specific social events or venues. For this, three main association steps are considered: \textit{geographical}, \textit{coverage-based} and \textit{performance-based}.
	
	\begin{figure*}[ht] 
		\centering
		\includegraphics[width=2\columnwidth]{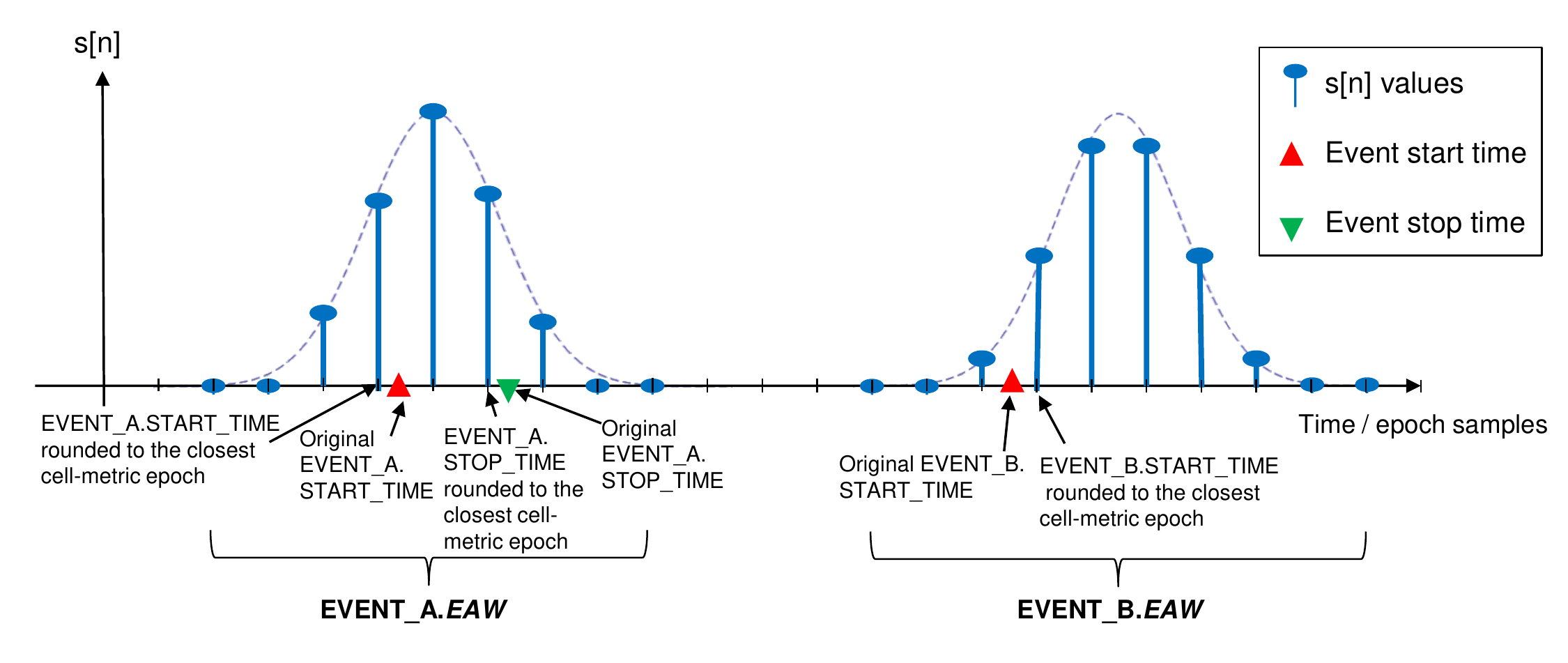}
		\caption{EAW and social indicator (s[n]) generation example.}
		\label{fig:Fig_3}
	\end{figure*}

	\subsubsection{Location and bearing}
	
	This submodule is used to generate a subset of NEs associated with each event according to their geographical location and orientation. 
	
	Firstly, for each event or venue, the sites standing within a defined maximum distance (MAX\_DIST) are considered as associated to the event. Additionally, a maximum and minimum number of associated sites (MAX\_NUM\_ASSOC\_SITES and MIN\_NUM\_ASSOC\_SITES, respectively) can be defined to both vary the distance limit and guarantee that a minimum number of sites is associated with each event. 
	
	Secondly, the angle between a point and the maximum of the radiation pattern of a cell is defined as the bearing between them. Low values of bearing indicate that the cell is oriented towards the element. In sectorial sites, the horizontal width of a cell (CELL.HOR\_WIDTH), referring to the width of the angular section to be covered by a cell, has also to be considered. For example, in tri-sectorial sites, CELL.HOR\_BW =120º, where bearings $\in$ [0º, 60º) would indicate locations in the cell angular field of coverage.
	
	In this way, the submodule provides an initial lightweight association that can be quickly calculated for city-wide analysis. As a result, the output of this step can already serve as an input to some OAM tasks (as it is represented in \Fref{fig:Fig_2} by the dashed arrow from the submodule to the ``associated network and social data'' output). For example, it provides a list of past and predicted future events (including their details, ranking and relevance) that are close and in the angle of each cell of a region. 
	
	Further submodules would be based on this initial association as the starting point of more detailed analysis. As they are more computationally expensive, it is expected that they will be called for specific case analyses, such as a particular network degradation.

	\subsubsection{Coverage-based}
	
	Radio maps, propagation models or geolocated UE traces can be used to estimate the specific cells that are most likely to serve each social location UEs. However, these inputs are not available for all areas. Their processing is computationally expensive and they are typically stored outside the common operations and support system (OSS), formats and tools. 
	
	\subsubsection{Performance-based}
	
	This step analyzes the impact of past social events in the network performance. This is of key applicability in two fields. On the one hand, for the analysis and troubleshooting of past network events and degradations. On the other hand, for the prediction of future impacts in the same NEs or by other events in the same venue or of the same type. 
	
	The events can be analyzed individually, however, when analyzing the impact of venues, all the events hosted in that venue should be included. Also, any specific set of the available network metrics could be used as inputs. Metrics showing periodical patterns (e.g., daily traffic) should be normalized in order to properly assess the impact of social events.
	
	For each event, an \textit{event association window} (EAW) is defined, identifying the epoch (or index) of the cellular metric to be considered, delimiting its temporal scope. Following the scheme in \Fref{fig:Fig_3}, the EAW will contain the epochs surrounding each of the events start time. 
	
	If the stop time of the event is available, the EAW would contain a certain number of samples after its rounded value. Conversely, if the stop time is not available, which is the most common case, the length of the EAW would be estimated based on expected duration. 
	
	Once the EAWs have been defined, different procedures can be applied for the association between the events and the cellular metrics. For example, this might be based on slope detection in the EAWs. However, to have a numerical feedback of the level of association, a metric-based comparison mechanism is adopted. This is supported by the proposed concept of \textit{social indicator}, consisting in a metric, s[n] (where n stands for the sample index), generated to model the evolution of the expected impact of events. The s[n] is defined to share the same temporal characteristics, such as the periodicity (hourly, daily), of the network metrics to be analyzed, but it is constructed based only on social event information.   

	In this way, dimensionless values of s[n] should follow the expected impact on the network metrics of a social event. From the analysis of the network behavior, the impact of an event increases as the start time gets closer and decreases towards its ending and beyond. To model this behavior, a baseline definition mechanism is established, consisting in assigning to the s[n] a normal-like distribution with mean equal to the middle of the EAW and a standard deviation, $\sigma$, defined to reflect the expected slope of the event impact. Other options such as Gaussian mixture models, could properly fit to some ``multi-peak'' events (e.g. those with internal periods of lesser attendance). However, given that s[n] is based on the social data and this typically does not include details on possible intermediate behavior (only start and end times), the multiple-peak nature of an event cannot be inferred from it. 
	
	The values of s[n] can then be compared to any cellular metric (like those automatically gathered by the OSS of the cellular network), represented as x[n] (e.g., the number of dropped calls) to find possible relations between events and the network performance. 
	
	To find these possible relations, the Pearson correlation coefficient ($r_{EAW}$) is adopted: a $|r_{EAW}|$ value close to 1 indicates high correlation between the event and the cellular metric. $|r_{EAW}|$ values close to 0 indicate little or no relation between s[n] and x[n]. The sign of $r_{EAW}$ indicates if the event is associated to an increase in x[n] (if $r_{EAW}>0$) or a decrease (if $r_{EAW}<0$). 
	
	For the analysis of venues having multiple events (and therefore, multiple EAWs), different approaches are possible. In our baseline application, the level of correlation is obtained for each of the multiple EAWs. Then, different statistical analysis can be done based on this information: the median of the impacts generated by the events taking place in the venue, their maximum, the relation between the day of the week of events and their impact, etc. 

	The calculation in different independent EAWs have also the advantage of determining their impact just based on their correlation. Prior scaling of the events based on expected attendance could be applicable, but they can be misleading as it also implies many additional factors (e.g. type of event, technological level of the crowd, etc.).

	\begin{figure*}[t!] 
		\centering
		\includegraphics[width=2\columnwidth]{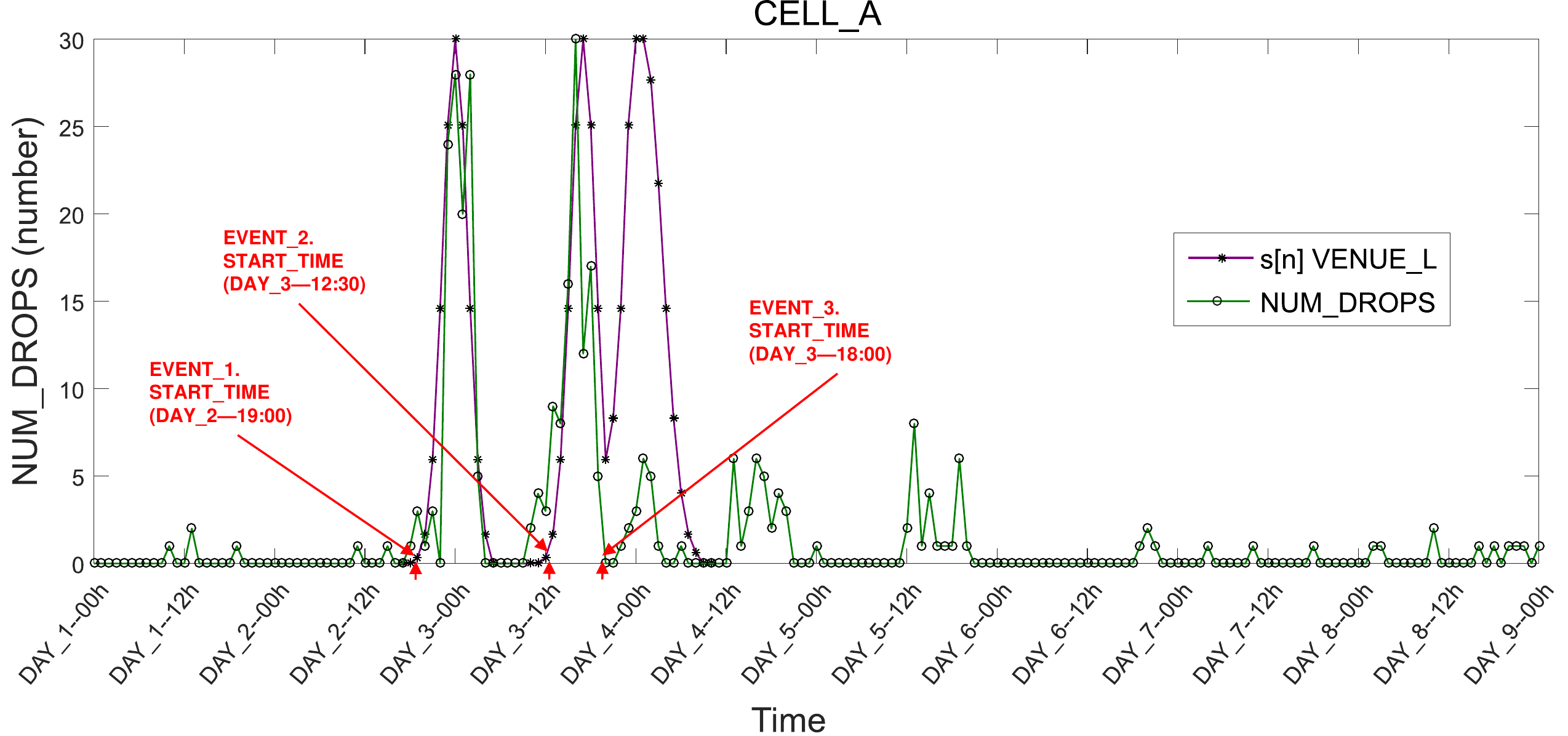}
		\caption{Hourly number of drops for CELL\_1A and the social indicator for VENUE\_L.}
		\label{fig:Fig_4}
	\end{figure*}

	\subsection{Outputs production}

	The outputs typically include the events and venues processed information consisting in the ranked list of events/venues with their different fields calculated during the process (NAME, START\_TIME, END\_TIME, VENUE, ADDRESS, LAT, LON, GEOGRAPHICAL\_CLOSE\_SITES, CORRELATED\_CELLS, etc.). 
	
	\section{Evaluation} \label{sec:Evaluation}

For the evaluation, the proposed framework was implemented in Matlab and executed in a PC. Cellular network data and metrics were gathered by query from replicated databases of those of the network OAM systems. The different blocks of the framework took advantage of the available functions for API calling to get information from social and consolidation sources. 
	
	The system was applied to the analysis of a LTE real cellular network based on macrocells (coverage areas in the range kilometers). This network has around 7500 cells and 850 sites (around 9 cells per site). The analysis is centered on the capital of the region, a city of around one million inhabitants. The process will be followed through the presented steps of the system.
	
	The evaluation in this case focuses on the application of the proposed scheme for the identification of possible social events causing degradation in the service provision. In this case, for all cells and events, all the processing steps will be calculated except for the performance-based network association (as it is the most time consuming one and it should be done periodically). This last step will be taken only if a degradation is detected in a cell/sector, in order to establish whether the degradation is due to a social event, finally leading to its identification.
	
	As input for the acquisition module, the geographical scope is provided in the form of the tuple [COUNTRY, REGION, CITY]. For the temporal scope, the input includes the initial and last date under consideration, covering a period of 54 days. For this period, module A acquires the social data from one unique social source, an online event calendar. This returns the parameters START\_TIME, LON, LAT, ADDRESS, VENUE and POPULARITY of 2200 events, where 600 distinct venues are identified.

	In this block, firstly, the events with empty geographical coordinates are discarded. Secondly, those with terms \{``bar'', ``cafe'', ``coffee'', ``pub'', ``tavern'', ``inn'', ``church'', ``shop'', ``club'', ``gospel'', ``lounge''\} in their venue name are eliminated by the semantic filtering as, for the particular region under analysis, these types of places are not expected to generate large crowds, in terms of being large enough to impact macrocell services. In any case, these considerations would change from one region to another and from time to time, not being directly applicable to all deployments. 
	
	Finally, addresses whose region is outside the REGION of interest are filtered out. This process leaves 1768 events, performed in 507 different venues.

	\subsection{Cellular network association}
	
	The initial location-based association between sites and each specific event is performed considering expert-defined loose restrictions MAX\_NUM\_CLOSE\_SITES = 7, MIN\_NUM\_CLOSE\_SITES = 1 and MAX\_DIST = 2 (kms). These parameters came from the experience with LTE macrocell not-dense environments, trying to gather all sites that might be affected by an event. Denser deployments would require smaller distances and more sites for a complete study. The obtained information is added to the filtered social data provided by the module B (\Fref{fig:Fig_2}), which includes the events details, the closest sites to each event and their distances. 
	
	\begin{figure}[ht] 
		\centering
		\includegraphics[width=1\columnwidth]{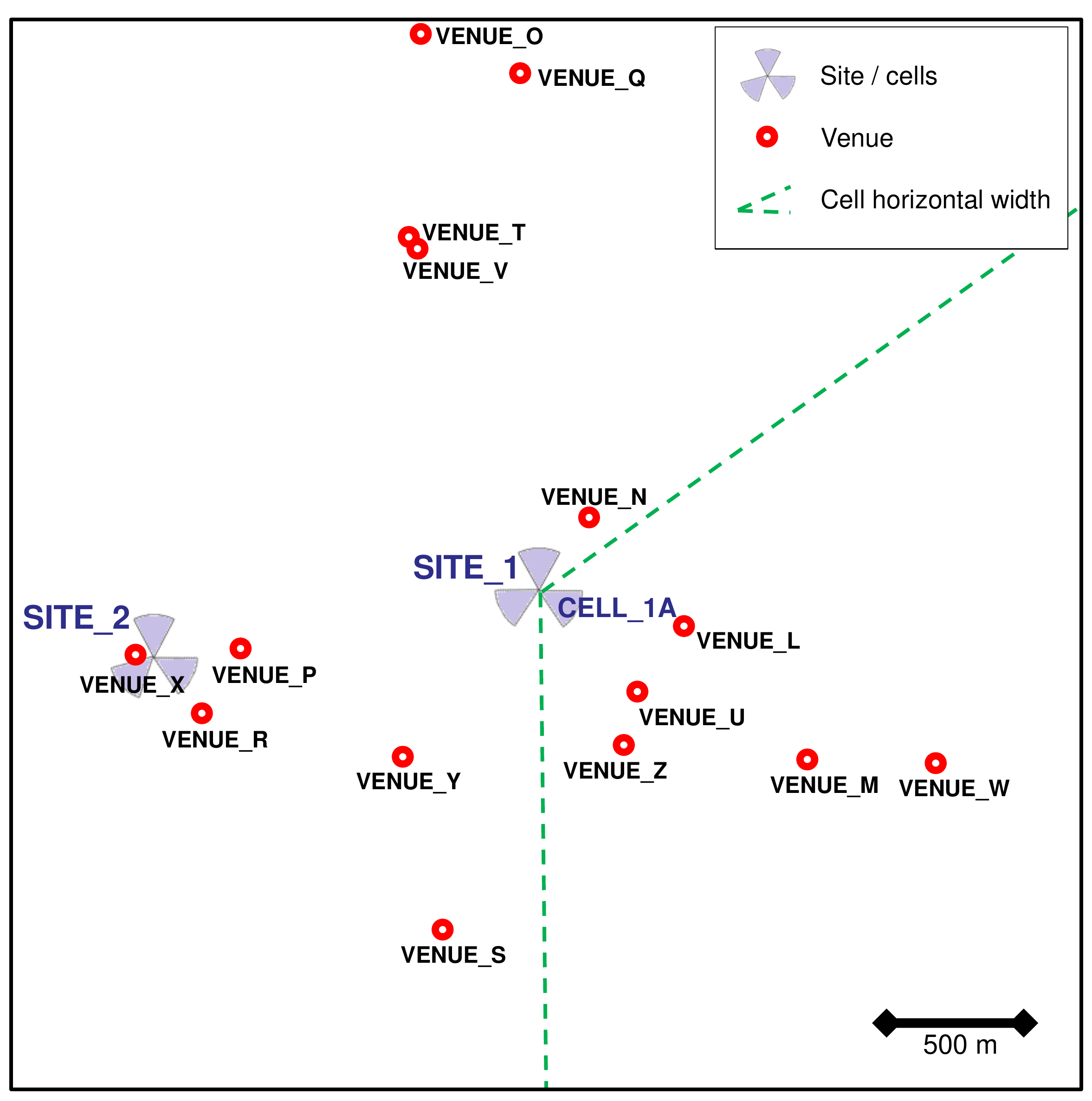}
		\caption{Filtering of CELL\_1A associated venues by location and bearing.}
		\label{fig:Fig_5}
	\end{figure}

	\iftrue
	\begin{table*}[ht] 
		\footnotesize
		\renewcommand{\arraystretch}{\asta}
		\caption{CELL\_1A associated venues analysis in module C.}
		\label{table:Filtering}
		\centering %
		\begin{tabular}{@{} c p{0.14\textwidth} c c c c c c @{}}
			\toprule[\tpw]   
			VENUE IDs	& TYPE &	DISTANCE 	&	BEARING	&	N\_EVENTS	&	NUM\_RRC\_CONN	&	NUM\_DROPS	&	DL\_USER\_THR	\\
			& &(kms)	&	(degrees)	&		&	$|r_{EAW}|$	&	$|r_{EAW}|$	&	$|r_{EAW}|$	\\
			\midrule
			VENUE\_L	& Large auditorium  &	0.56	&	15.22	&	3	&	0.83	&	0.73	&	0.84	\\
			VENUE\_M	& Religious center &	1.2	&	4.46	&	1	&	0.13	&	0.19	&	0.26	\\
			VENUE\_U	& Events pavilion &	0.53	&	18.63	&	1	&	0.24	&	0.03	&	0.11	\\
			VENUE\_W	& Private shop &	1.61	&	5.79	&	1	&	0.07	&	0.07	&	0.03	\\
			VENUE\_Z	& Exhibit hall&	0.67	&	33.17	&	1	&	0.28	&	0.4	&	0.12	\\
			\bottomrule
		\end{tabular}
	\end{table*}
	\fi

	\subsection{Application for the identification of social events - based degradations}
	
	At this point, the system contains a detailed list of the events, venues and the closest sites for each of them. This information could be already passed to module D (outputs productions) and used as input for other multiple OAM tasks in the cellular network, both manual or automatic. 
	
	The last block of the proposed system, the performance-based analysis, is assessed for a specific application case: the automatic identification of the social event that could be causing a degradation. In particular, the sector CELL\_1A, located in the site SITE\_1, is analyzed. This cell presents a degradation that was discovered by a simple threshold-based detection mechanism. The degradation consists in two important increases in the number of drops (NUM\_DROPS) for some hours in two consecutive dates, as it is presented in \Fref{fig:Fig_4}. This degradation can be generated by multiple reason and it was identified by the automatic OAM tools monitoring the area.
	
	To investigate the problem, the available social information is analyzed. As input for the following steps the events that include SITE\_1 in the list of close sites are passed to the next stage. This leaves 29 events in the 15 venues shown in \Fref{fig:Fig_5}. 
	
	For these venues, bearing filtering is applied considering CELL\_1A.AZIMUTH (120º) and CELL\_1A.HOR\_WIDTH (120º). \Fref{table:Filtering} shows those venues in \Fref{fig:Fig_5} that passed the filtering (bearing $<$ 60º), including their distances and bearing. Also the type of venue is included as reference. Here, it has to be noticed that the system does not take into account this information, focusing only in the geographical, coverage and performance-based association with the network data.

	Then, the correlation is obtained for the filtered events and a set of CELL\_1A KPIs: the number of connections (NUM\_RRC\_CONN), the number of drops (NUM\_DROPS) and the downlink throughput (DL\_USER\_THR). These KPIs are selected as they are commonly related to peaks of service and the analyzed drop degradations.
	
	The social indicator, s[n], is generated considering the normal-based traffic profile behavior. The correlations results are also included in \Fref{table:Filtering}, showing the $|r_{EAW}|$ mean values for the events of each venue. Here, the column N\_EVENTS indicate the number of events in each venue for the DAY\_1 to DAY\_9 period (the range of dates where KPIs were available). It can be observed how VENUE\_L is the only one whose KPIs present high correlation values ($|r_{EAW}|>0.7$).
	
	Analyzing the three events of VENUE\_L individually (see \Fref{fig:Fig_4}), it is observed how its first two (EVENT\_1 and EVENT\_2) fit perfectly with the identified two peaks in the number of drops. Taking this information into account, the system used for troubleshooting can directly identify EVENT\_1 and EVENT\_2 as those causing the degradation. EVENT\_3 also presents important correlation values, although it is associated with a smaller degradation happening after the first two. In fact, EVENT\_3 it was a very minor sport event, meanwhile EVENT\_1 and EVENT\_2 were a massive concert and a big political meeting, respectively. The fact that the proposed system is able to provide feedback at this level, without human intervention is a good indication of the capabilities of the proposed approach.
	
	From this, it can be expected that future events in VENUE\_L would produce similar effects and impact on the network. Here, a possible recommendation could be the preallocation of resources for those events.

	\section{Conclusions and Outlook}\label{sec:Conclusion}
	
	This work has presented the main use cases and general scheme for the application of social data information acquired from event databases and calendars to the management of cellular networks. A complete framework and detailed steps for the association of such information with cellular network details and metric is established. This is applied in a real field use case in the identification of events behind a cause of failure. 
	
	The results show the presented approach as a powerful tool to support varied OAM and SON tasks. In this way, the increasing availability of social data sources and the advances of social analysis, data fusion and big data techniques are expected to be increasingly integrated in the management activities and cellular networks allowing to a better fit between the network services and the changing demand of the users.  
	
	\section*{Acknowledgments}
	
	This work has been partially funded by Optimi-Ericsson, Junta de Andalucía (Agencia IDEA, Consejería de Ciencia, Innovación y Empresa, ref.59288, and Proyecto de Investigacion de Excelencia P12-TIC-2905) and ERDF. This work has been also partially performed in the framework of the Horizon 2020 project ONE5G (ICT-760809) receiving funds from the European Union. The authors would like to acknowledge the contributions of their colleagues in the project, although the views expressed in this contribution are those of the authors and do not necessarily represent the project.

	\bibliographystyle{IEEEtran}
	\bibliography{./BIBTEX/Bibliography} 

	\ifCLASSOPTIONcaptionsoff
	\newpage
	\fi
	
	\iftrue
	\section*{Author Information}
	
	\begin{IEEEbiographynophoto}{SERGIO FORTES (sfr@ic.uma.es)} holds a M.Sc. and a Ph.D. in Telecommunication Engineering from the University of Málaga. He began his career being part of main european space agencies (DLR, CNES, ESA) and Avanti Communications plc, where he participated in various research and consultant activities on broadband and aeronautical satellite communications. In 2012, he joined the University of Malaga, where his research is focused on Self-Organizing Networks for cellular communications. 
	\end{IEEEbiographynophoto}
	
	\begin{IEEEbiographynophoto}{DAVID PALACIOS (dpc@ic.uma.es)} received the M.Sc. degree in telecommunication engineering from the Universidad de Málaga, Málaga, Spain, in 2013, where he is currently pursuing the Ph.D. degree. From 2013, he has worked as a Research Assistant with the Department of Communications Engineering, Universidad de Málaga, in the development of SON (Self-Organizing Networks) functionalities for automatic fault detection and self-optimization in cellular networks. He is also the recipient of a Junta de Andalucía Scholarship (2016-2019).
	\end{IEEEbiographynophoto}

	\begin{IEEEbiographynophoto}{RAQUEL BARCO (rbm@ic.uma.es)} holds a M.Sc. and a Ph.D. in Telecommunication Engineering from the University of Malaga. From 1997 to 2000, she worked at Telefónica in Madrid (Spain) and at the European Space Agency (ESA) in Darmstadt (Germany). In 2000, she joined the University of Malaga, where she is currently Associate Professor. She has worked in project with the main mobile communications operators and vendors and she is author of more than 100 high impact journals and conferences.
		
	\end{IEEEbiographynophoto}

	\begin{IEEEbiographynophoto}{INMACULADA SERRANO (inmaculada.serrano@ericsson.com)} received the MSc degree from the Universidad Politécnica Valencia. She specialized further in radio after complementing her education with the master's degree in mobile communications. She joined Optimi and started a wide career in the optimization and troubleshooting of mobile networks, including a variety of consulting, training and technical project management roles, in 2004. She moved to the Advanced Research Department, Ericsson, in 2012.
	\end{IEEEbiographynophoto}

	\fi


\end{document}